\documentclass[10pt,english]{article}

\usepackage{mathptmx}
\usepackage[T1]{fontenc}
\usepackage[utf8]{inputenc}
\usepackage{geometry}
\usepackage{color}
\usepackage{babel}
\usepackage{float}
\usepackage{amsmath}
\usepackage{amsthm}
\usepackage{graphicx}
\usepackage{esint}
\usepackage{amsfonts}
\usepackage{amssymb}
\usepackage{lineno}
\usepackage{stackrel}
\usepackage{units}
\usepackage{mathrsfs}
\usepackage{setspace}
\usepackage{multicol}
\usepackage{textcomp}

\usepackage[super,sort&compress,numbers]{natbib} 
\usepackage{soul}
\usepackage[unicode=true,pdfusetitle,
 bookmarks=true,bookmarksnumbered=true,bookmarksopen=true,bookmarksopenlevel=13,
 breaklinks=false,pdfborder={0 0 0},pdfborderstyle={},backref=false,colorlinks=true]
 {hyperref}

\usepackage{url}
\usepackage[export]{adjustbox}[2011/08/13]

\renewcommand{\cite}[1]{\citep{#1}} 

\makeatletter
\newcommand{\lyxaddress}[1]{
	\par {\raggedright #1
	\vspace{1.4em}
	\noindent\par}
}

\@ifundefined{date}{}{\date{}}
\usepackage{color}
\usepackage[usenames,dvipsnames,svgnames,table]{xcolor}
\usepackage{multicol}
\definecolor{burgundy}{rgb}{0.5, 0.0, 0.13}
\definecolor{airforceblue}{rgb}{0.36, 0.54, 0.66}
\hypersetup{urlcolor=airforceblue}
\hypersetup{citecolor=burgundy}

\makeatother
\usepackage{etoolbox}
\begin{document}
\title{The deep atmosphere of Jupiter}
\author{\href{https://orcid.org/0000-0003-4428-2700}{Keren Duer-Milner}$^{1,2\star}$}
\maketitle

\lyxaddress{\begin{center}
\textit{$^{1}$Leiden Observatory, Leiden University, Niels Bohrweg
2, NL-2333 CA Leiden, the Netherlands.}\\
\textit{$^{2}$SRON Netherlands Institute for Space Research, Niels
Bohrweg 4, 2333 CA, Leiden, the Netherlands.}\\
\small$^\star$Corresponding author. Email: duer@strw.leidenuniv.nl\\
\par\end{center}}

\lyxaddress{\begin{center}
Published as a Comment in Nature Communications\textit{}\\
\par\end{center}}

\begin{abstract}
    \textbf{Jupiter, a rapidly rotating gas giant, features over 20 atmospheric jet streams that penetrate thousands of kilometers into the planet. This work discusses recent progress, identifies key uncertainties regarding the jets' driving and dissipating mechanisms, and suggests future research avenues. }
\end{abstract}
\vspace{3mm}
\begin{multicols}{2}

Jupiter, the solar system's largest planet, has long captivated astronomers. Its rapid rotation (approximately 10 hours), colorful bands, and turbulent weather layer—exhibited, for example, in the Great Red Spot—indicate a dynamic atmosphere that has been active for centuries. Unlike Earth, Jupiter lacks a solid surface, and its largely homogeneous composition simplifies some aspects of atmospheric study. However, the processes driving its atmospheric features are still largely a mystery. Recent research, spurred by data from NASA's Juno mission and state-of-the-art computer simulations, is beginning to reveal the secrets of Jupiter's atmosphere, particularly its prominent zonal jets—high-speed winds that encircle the planet in the east-west direction.

\begin{figure*}
    \centering
    \includegraphics[width=0.8\linewidth]{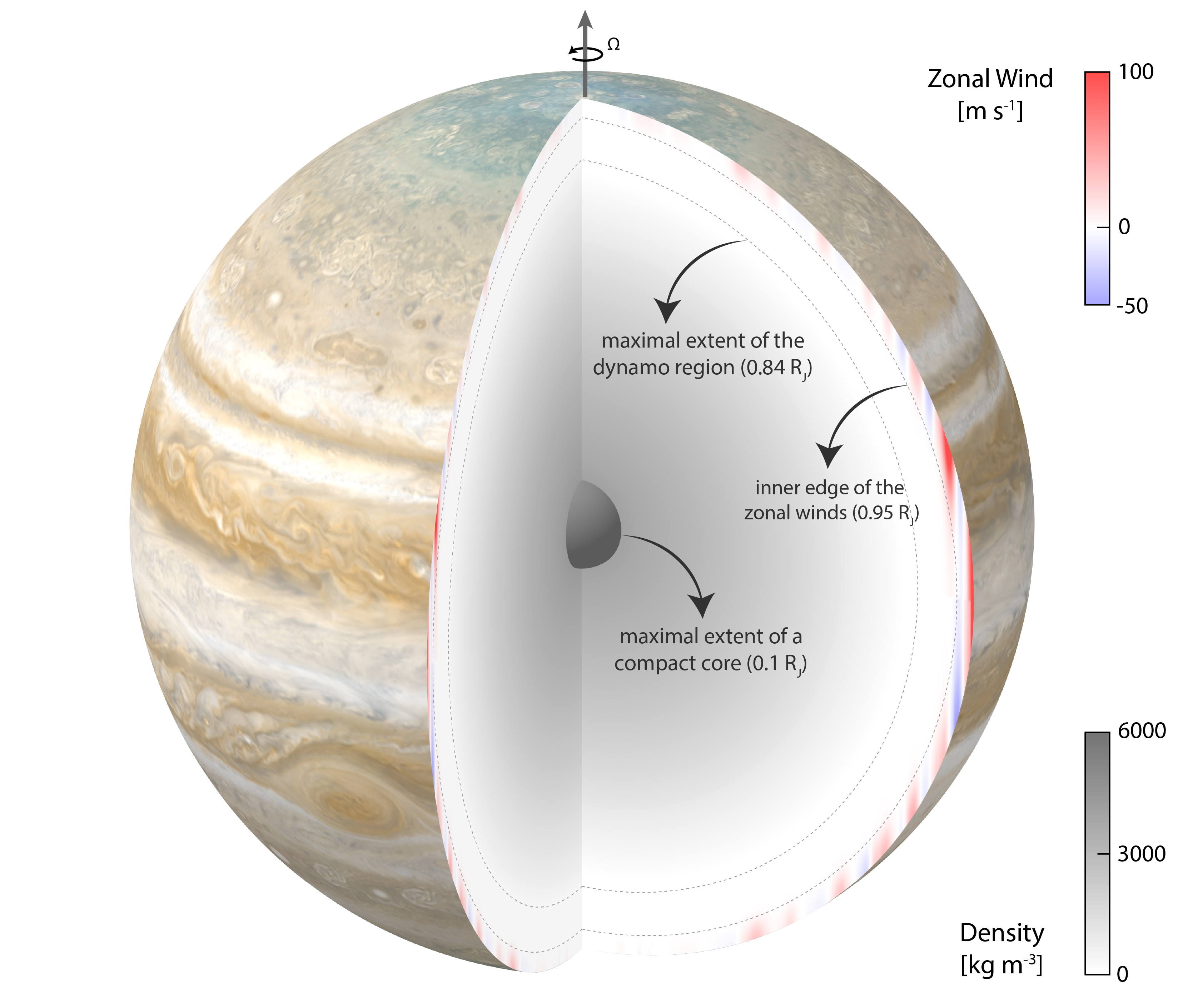}
    \caption{Graphic representation of Jupiter's winds and internal structure. The figure depicts a three-dimensional representation of the planet, showing the zonal winds (eastward in red, westward in blue) and a grayscale shading that illustrates a potential density structure of the interior. Arrows indicate the maximal edge of the dynamo region ($0.84\:R_{\rm{J}}$, where $R_{\rm{J}}$ is Jupiter's radius), the inner edge (penetration depth) of the zonal winds ($0.95\:R_{\rm{J}}$), and the maximal extent of a compact core ($0.1\:R_{\rm{J}}$). Jupiter's clouds image credit: 'Merged Cassini and Juno global map of Jupiter' by NASA/JPL-Caltech/SSI/Southwest Research Institute/Malin Space Science Systems/Italian Space Agency (ASI)/Italian National Institute for Astrophysics (INAF)/JIRAM/Bj\"orn J\'onsson, licensed under CC BY 3.0. No changes were made to the original image.}
    \label{fig: 3D fig}
\end{figure*}

These jet streams are particularly fascinating from an atmospheric dynamics perspective. Wind velocities reach approximately 100 $\rm{ms^{-1}}$, more than three times the fastest jet stream on Earth. The wind direction alternates between east and west, with over 20 distinct jet streams observed \cite{Ingersoll1990,sanchez2023}. This pattern, where winds alternate in direction with latitude, is quite unlike Earth's more structured jet system. Notably, at Jupiter's equator, the wind flows in the direction of the planet's rotation. {This phenomenon, known as 'superrotation,' requires momentum to be transported toward the equator, something that the planet's rotation alone cannot accomplish \cite{imamura2020}.} An illustration of these winds is available in Figure~\ref{fig: 3D fig}, where eastward (westward) winds are shown in red (blue) color. Measurements of Jupiter's wind speeds have been conducted over centuries, from early observations since Galileo’s time
using small telescopes, through space missions like Voyager and
Cassini, to large current-day telescopes like Hubble and JWST.

\section*{Measuring dynamics in the Jovian Atmosphere}

The most direct method to measure the deep features of a planetary atmosphere is to send a probe to obtain \textit{in situ} measurements. This was achieved for Jupiter only once, when the Galileo entry probe descended into Jupiter's atmosphere in 1995. It measured super-solar abundances of heavy elements, the abundance of Nobel gas (which determined the presence of helium rain in the planet), and an increase in wind speed compared to cloud-level values until a pressure of 20 bars (Note: 1 bar is approximately the atmospheric pressure at Earth's surface) \cite{Atkinson1998}. However, later studies suggested that the Galileo entry point was a local 'hotspot' and did not necessarily reflect a global trend \cite{Showman2000}.

Comprehensive measurements of the two-dimensional wind velocity field at the cloud level ($\sim1$ bar) were provided by the Cassini spacecraft. During its Jupiter flyby in December 2000, Cassini captured consecutive images of the atmosphere, allowing for the calculation of wind velocity by tracking cloud movements  \cite{Salyk2006,Choi2011,galperin2014cassini}. These measurements enabled comparison of the jets' wind velocity with turbulent fluxes, revealing a strong correlation. This suggested that Jupiter's jet streams are, at least in part, eddy-driven, analogous to one of the mechanisms operating in Earth's atmosphere, where eddies (turbulence) transport momentum upscale to the zonal jets. However, information on the atmosphere beneath the cloud layer remained scarce until recently.

A significant advancement comes from Juno's mapping of Jupiter's gravity field \cite{Bolton2017,Iess2018}. By precisely measuring subtle variations in Jupiter's gravitational pull on the Juno spacecraft, details about the planet's internal structure, mass distribution, and atmospheric properties have been revealed. The low-order gravity harmonics indicate that Jupiter's envelope is inhomogeneous, and a compact core is plausible but much smaller than previously assumed \cite{Wahl2017,miguel2022}. An illustration of such an envelope is apparent in grayscale in Figure~\ref{fig: 3D fig}. {The profile shown here is the one with the \textit{largest} possible compact core (dark gray)\cite{ziv2024characterizing,howard2023jupiter,miguel2022}; however, it is not yet determined whether Jupiter possesses such a compact core in addition to a dilute one. In all recent models, a dilute core (lighter gray, $\sim0.1-0.5 R_{\rm{J}}$), where heavy elements are spread throughout the planet's envelope, is essential for accurately explaining its gravitational measurements \cite{HELLED202651}. The asymmetric gravity field, and the higher-order gravitational harmonics, has revealed that the zonal jets penetrate thousands of kilometers into the planet's interior \cite{Kaspi2018}, where the pressure level reaches approximately $10^5$ bars (shown in red-blue colors in Figure~\ref{fig: 3D fig}). This discovery solved a long-standing question about jets' structure in the Jovian atmosphere. Below the jet-streams, Jupiter rotates like a solid body \cite{Guillot2018}.}


Later studies revealed that the majority of the gravity signal originates from the jets at approximately $\pm20$ degrees latitude \cite{galanti2021Con}. These jets must penetrate the interior cylindrically \cite{kaspi2023}, emphasizing the importance of rotation in Jupiter's atmosphere, where the jets are aligned with surfaces of constant angular momentum.

Intriguingly, this deep penetration is consistent with independent estimates derived from a phenomenon called ohmic dissipation \cite{Liu2008}. As the electrically conductive gas within Jupiter moves through the planet's strong magnetic field, it generates electrical currents that dissipate energy in the form of heat. By comparing this process with the amount of heat emitted by Jupiter, a constraint on the depth of the zonal jets was estimated. At approximately $10^5$ bars, the electrical conductivity becomes significant, raising the possibility that the jets and the magnetic field are interacting \cite{Cao2017}. {This process is independent of the dynamo-generating region of the planet, which is expected to occur at deeper levels, extending up to Jupiter's Lowes radius ($0.84 \: R_{\rm{J}}$, see Figure \ref{fig: 3D fig}) \cite{sharan2022internal}.} Measurements of the time-varying magnetic field have provided no clear evidence for such an interaction \cite{connerney2022new,bloxham2024rapidly}, challenging our understanding of the deep mechanisms that maintain Jupiter's jet streams against dissipation.

\section*{The Puzzling Atmospheric Dynamics}

The processes limiting or "quenching" jet strength at depth in gas giants are not well understood. These include the effects of magnetic drag \cite{Liu2010}, where the magnetic field acts like a brake on the winds; large density variations; and localized stable stratification, where layers of different density resist mixing and suppress convection \cite{christensen2024quenching}. Current theoretical and numerical frameworks favor stable layers as an efficient dissipating mechanism \cite{wulff2022}, but the origin of such a layer remains a mystery. Some proposed scenarios include a "helium rain" layer, estimated to exist at pressures of $10^6$ bars or deeper \cite{brygoo2021evidence} (a region also believed to generate the planetary dynamo, see Fig. \ref{fig: 3D fig}), and a 'deep' radiative zone \cite{Guillot1994,siebenaler2025conditions}, estimated at $10^3-10^4$ bars. However, these have been challenged by observations of the jets' extent, leaving this question open for future exploration.

Mechanisms that provide energy to the jets, often termed "jet-pumping" mechanisms, differ for the equatorial and higher-latitude jets. The higher-latitude jets, poleward of approximately $\pm 17$ degrees latitude, are believed to be powered by atmospheric turbulence \cite{Choi2011,young2017}. It is still uncertain whether this turbulence is confined to a shallow weather layer, like Earth's baroclinic eddies, or originates from deep convective plumes \cite{Heimpel2016}. Both shallow weather layer Global Circulation Models (GCMs) powered by solar fluxes \cite{Liu2010} and deep convective GCMs powered by internal heat \cite{duer2023} show that the jet streams and small-scale turbulence are interdependent.

{The equatorial jet, as mentioned earlier, requires momentum transport to the equator to sustain an eastward jet stream.} Such a source has multiple possible origins, including water condensation releasing latent heat \cite{Lian2010}, north-south wave convergences \cite{Liu2010}, parameterized convection \cite{showman2019}, and heat transport from the interior by organized convection \cite{Busse1994,heimpel2022}. Recent evidence points to the existence of these deep heat fluxes \cite{duer2024}; however, precisely characterizing the jet-pumping mechanism requires further research and measurements.

Overturning circulation, the north-south and up-down components of the flow field, also plays a role in shaping Jupiter's atmosphere. Data from the Microwave Radiometer (MWR) instrument on Juno suggest that Jupiter's atmosphere at midlatitudes is organized into a series of alternating circulation cells, similar to Earth's Ferrel cells but much deeper \cite{duer2021}. These cells, characterized by rising and sinking air masses, could play a key role in transporting heat and momentum throughout the atmosphere. As these cells are aligned with turbulence, it has been suggested that they are also driven by turbulent fluxes in the atmosphere. This currently represents the only available evidence for deep overturning circulation on a giant planet.

Interestingly, the overturning circulation above the cloud layer, where the atmosphere is expected to be stably stratified, suggests a reverse circulation pattern \cite{fletcher2020}. These measurements are derived from temperature assessments based on visible and infrared observations. This raises the possibility of stacked meridional circulation, where an eastward jet in the northern hemisphere exhibits counterclockwise circulation beneath the clouds and clockwise circulation above them. Such a circulation pattern is not found in Earth's atmosphere and is unique in numerical simulations, which require a frictional mechanism above the clouds to produce such a pattern \cite{duer2023}. The meridional circulation in the equatorial region indicates upwelling \cite{Li2017}, but the mechanism creating such an upwelling is still an open question. It has been suggested that organized convection can lead to such massive upwelling \cite{duer2021}, strengthening the possibility that this mechanism is active in Jupiter's equatorial region.

\section*{The future of Jupiter's atmospheric exploration}

Despite the remarkable progress in recent years, many fundamental questions regarding Jupiter's atmospheric dynamics remain unanswered. Future research must continue to push the boundaries of both observational capabilities and computational modeling.

On the observational front, long-term monitoring of Jupiter's atmosphere, spanning from visible to microwave and infrared wavelengths, is crucial for capturing the temporal variability of zonal jets, vortices, and potentially even the overturning circulation. Future space missions could deploy dedicated atmospheric probes to multiple locations, providing detailed vertical profiles of temperature, wind velocity, and chemical composition to greater depths than currently available. Such measurements are essential for confirming the mechanisms driving Jupiter's atmospheric dynamics.

Furthermore, exploration of Jupiter's magnetic field and its interaction with the deep atmosphere warrants further investigation. The second extended mission of Juno might allow reanalysis of potential jet-induced currents, as the extended timescale (exceeding 10 years) will enable proper time variation analyses \cite{wicht2024contributions}.

On the modeling front, the next generation of GCMs must incorporate increasingly realistic physical processes, including sophisticated treatments of radiative transfer, cloud microphysics, and deep internal heating within the same model. This poses a significant challenge, as modeling a 'deep' atmosphere often precludes resolving the upper observable layers of Jupiter. Improvements in equations of state and immiscibility curves (for H-He demixing) are crucial for understanding the chemistry of the deep atmosphere and the locations of phase transitions, as these will directly translate into the depth and structure of the convective layer.

Ultimately, unraveling the secrets of Jupiter's atmosphere requires a synergistic approach, combining cutting-edge observations with state-of-the-art numerical simulations. By addressing the key questions outlined above, we can hope to achieve a truly comprehensive understanding of the forces shaping the dynamics of the solar system's largest planet.

\section*{Open Research}
No new data sets were generated during the current study. The density profile in Figure \ref{fig: 3D fig} is available in \cite{ziv2024characterizing}, and the wind profile is available in \cite{kaspi2023}. 

\subsection*{Acknowledgments}
I thank Yamila Miguel, Nimrod Gavriel, Emily Sandford, Solène Ulmer-Moll, and Louis Siebenaler for interesting discussions, Maayan Ziv and Saburo Howard for providing interior profiles \cite{ziv2024characterizing,howard2023jupiter}, and the anonymous reviewer for the constructive feedback. I also thank the Institute for Environmental Sustainability (IES) at the Weizmann Institute of Science for the Next-gen Environmental Sustainability Postdoc award and the Council for Higher Education in Israel for the CHE/PBC Fellowship for Postdoctoral training abroad for women for providing personal financial support. This work is funded by the European Research Council (ERC) under the European Union’s Horizon 2020 research and innovation programme (grant agreement no. 101088557, N-GINE).

\subsection*{Competing Interests}
Authors declare that they have no competing interests.

\bibliographystyle{naturemag}
\bibliography{kerensbib}

\end{multicols}
\end{document}